\documentclass[journal=jpccck, manuscript=article, format=reprint, layout=onecolumn]{achemso}

\usepackage[version=3]{mhchem} 
\usepackage{xcolor}
\usepackage{booktabs}
\usepackage{natbib}
\usepackage{float}

\author{Mikhail E. Zaytsev}
\affiliation{Physics of Fluids, Max Planck Center Twente for Complex Fluid Dynamics and J.M. Burgers Centre for Fluid Mechanics, MESA+ Institute for Nanotechnology, University of Twente, P.O. Box 217, 7500AE Enschede, The Netherlands}
\alsoaffiliation{Physics of Interfaces and Nanomaterials, MESA+ Institute for Nanotechnology, University of Twente, 7500AE Enschede, The Netherlands}

\author{Yuliang Wang}
\affiliation{Robotics Institute, School of Mechanical Engineering and Automation, Beihang University, Beijing 100191, P.R. China}

\author{Yuhang Zhang}
\affiliation{Department of Mechanical Engineering, Johns Hopkins University, Baltimore, Maryland 21218, USA}

\author{Guillaume Lajoinie}
\affiliation{Physics of Fluids, Max Planck Center Twente for Complex Fluid Dynamics and J.M. Burgers Centre for Fluid Mechanics, MESA+ Institute for Nanotechnology, University of Twente, P.O. Box 217, 7500AE Enschede, The Netherlands}
\alsoaffiliation{TechMed Centre, University of Twente, Enschede, The Netherlands}

\author{Xuehua Zhang}
\affiliation{Department of Chemical and Materials Engineering, University of Alberta, 12-211 Donadeo Innovation Centre for Engineering, Edmonton, Alberta, Canada}
\alsoaffiliation{Physics of Fluids, Max Planck Center Twente for Complex Fluid Dynamics and J.M. Burgers Centre for Fluid Mechanics, MESA+ Institute for Nanotechnology, University of Twente, P.O. Box 217, 7500AE Enschede, The Netherlands}

\author{Andrea Prosperetti}
\affiliation{Physics of Fluids, Max Planck Center Twente for Complex Fluid Dynamics and J.M. Burgers Centre for Fluid Mechanics, MESA+ Institute for Nanotechnology, University of Twente, P.O. Box 217, 7500AE Enschede, The Netherlands}
\alsoaffiliation{Department of Mechanical Engineering, Johns Hopkins University, Baltimore, Maryland 21218, USA}

\author{Harold J. W. Zandvliet}
\affiliation{Physics of Interfaces and Nanomaterials, MESA+ Institute for Nanotechnology, University of Twente, 7500AE Enschede, The Netherlands}
\email {h.j.w.zandvliet@utwente.nl}

\author{Detlef Lohse}
\affiliation{Physics of Fluids, Max Planck Center Twente for Complex Fluid Dynamics and J.M. Burgers Centre for Fluid Mechanics, MESA+ Institute for Nanotechnology, University of Twente, P.O. Box 217, 7500AE Enschede, The Netherlands}
\alsoaffiliation{Max Planck Institute for Dynamics and Self-Organization, Am Fassberg 17, 37077 G\"ottingen, Germany}
\email{d.lohse@utwente.nl}

\title{Gas-Vapor Interplay in Plasmonic Bubble Shrinkage}

\begin{document}




\begin{abstract}

The understanding of the shrinkage dynamics of plasmonic bubbles formed around metallic nanoparticles immersed in liquid and irradiated by a resonant light source is crucial for the usage of these bubbles in numerous applications. In this paper we experimentally show and theoretically explain that a plasmonic bubble during its shrinkage undergoes two different phases: first, a rapid partial bubble shrinkage governed by vapor condensation and, second, a slow diffusion-controlled bubble dissolution. The history of the bubble formation plays an important role in the shrinkage dynamics during the first phase, as it determines the gas-vapor ratio in the bubble composition. Higher laser powers lead to more vaporous bubbles, while longer pulses and higher dissolved air concentrations lead to more gaseous bubbles. The dynamics of the second phase barely depends on the history of bubble formation, i.e. laser power and pulse duration, but strongly on the dissolved air concentration, which defines the concentration gradient at the bubble interface. Finally, for the bubble dissolution in the second phase, with decreasing dissolved air concentration, we observe a gradual transition from a $R(t) \propto (t_0 - t) ^{1/3}$ scaling law to a $R(t) \propto (t_0 - t) ^{1/2}$ scaling law, where $t_0$ is the lifetime of the bubble and theoretically explain this transition.

\end{abstract}

\section{Introduction}

Plasmonic nanoparticles immersed in a liquid and irradiated by a resonant laser can rapidly heat up to high temperatures, due to their peculiar absorption ability, and transfer the heat to the surrounding liquid, causing its heating and subsequent evaporation. The dynamics of these plasmonic bubbles has been extensively studied at the nanoscale \cite{kotaidis2006,  hleb2010, hleb2014, lombard2014, katayama2014, hou2015, lombard2016, nakajima2016, maheshwari2018} as well as at the microscale\cite{richardson2009, liu2010, baffou2014jpc, baral2014, liu2015, chen2017, zaytsev2018}, due to numerous important implications in biomedical therapy \cite{lapotko2009, emelianov2009, baffou2013, shao2015, liu2014, fan2014}, nano/micro manipulation \cite{krishnan2009, zhang2011, zhao2014, tantussi2018, xie2017}, and enhancement of chemical reactions \cite{baffou2014,adleman2009}. 

It is important to distinguish between plasmonic nano- and microbubbles. Former are induced around individual nanoparticles by ultra-short highly focused laser pulses. These plasmonic nanobubbles have relatively short lifetimes and sizes in the nanoscale domain. \cite{hleb2010, hleb2014, lombard2014, hou2015, lombard2016} Plasmonic microbubbles are formed by irradiating an ensemble, \textit{e.g.} a two-dimensional array, of nanoparticles with a laser. The plasmonic microbubbles are not only larger, but they also live substantially longer. \cite{liu2010, baffou2014jpc, liu2015, zaytsev2018}
Recently, it has been shown that plasmonic microbubbles forming on the water-immersed gold nanoparticle arrays under resonant cw-laser irradiation undergo four different phases: first -- nucleation (and possibly coalescence), growth and collapse of a giant vapor bubble, second -- an oscillating bubble phase, third -- bubble growth, which is mainly governed by water vaporization and fourth -- slow bubble growth due to air diffusion from the bulk liquid \cite{wang2017, wang2018}. So, the interplay between gas and vapor is crucial for the dynamics of plasmonic microbubbles.

\begin{figure*}[h]
 	\includegraphics[width=0.9\textwidth]{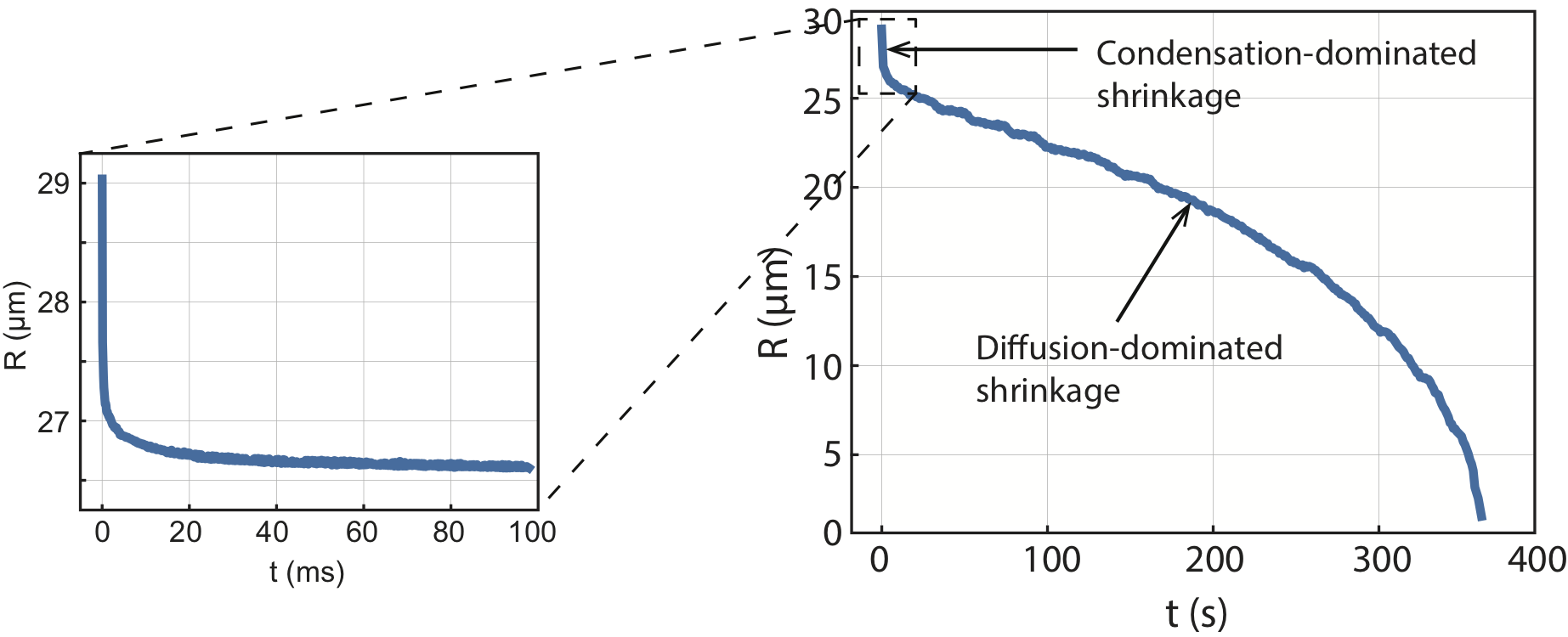}
 	\caption{
   The two different regimes are clearly distinguishable during plasmonic bubble shrinkage: phase 1 -- rapid bubble shrinkage governed by vapour condensation and phase 2 -- slow bubble dissolution due to air diffusion from the bubble to the bulk liquid.
 	}
 	\label{fig:bubble shrinkage scheme}
 \end{figure*}

After switching off the laser a plasmonic microbubble formed in water does not collapse immediately, but remains present on the substrate for minutes or even hours \cite{baffou2014jpc, liu2015}. That implies that the plasmonic bubble initially partially consists of air and, therefore the bubble is quite stable in the liquid. The process of this slow bubble dissolution is governed by air diffusion from highly-saturated areas in the vicinity of the bubble to the less saturated bulk liquid. Right before the process of slow bubble dissolution, rapid partial bubble shrinkage occurs, during the first milliseconds after switching off the laser, due to the fast vapor condensation. The pure vapor bubble collapse has been thoroughly scrutinized in numerous studies \cite{chao1965, plesset1977, brennen1995, legendre1998, hao1999, hao2000, prosperetti2017, pavlov2017}, while the collapse of the bubble, which consists of gas/vapor mixture is still a topic of current research \cite{nigmatulin1976, nigmatulin1981, hao2017, prosperetti2017}. A typical, experimentally obtained, dependence of the plasmonic bubble radius on time during shrinkage is shown in Figure \ref{fig:bubble shrinkage scheme}.

In our experiments the influence of various parameters, such as dissolved air concentration, which can be expressed in terms of gas saturation level $c_g/c_{sat}$, where $c_g$ is the air concentration and $c_{sat}$ the saturation air concentration, laser power $P_l$, and laser pulse duration $\tau_p$ during bubble growth on the subsequent bubble shrinkage is studied. The dynamics of the first phase (condensation-dominated) strongly depends both on the history of bubble formation, \textit{i.e.} laser power and laser pulse length, and on dissolved air concentration in the water, as they determine the bubble composition right before stopping of the laser irradiation. The second phase (diffusion-dominated) barely depends on the history of bubble formation, but strongly depends on the dissolved air concentration. The dissolved air concentration determines the effective scaling exponent for the bubble radius evolution in time. Our results bear important implications for the design and control of plasmonic microbubbles for various applications.

\section{Methods}
\subsection*{Sample preparation}
\begin{figure*}[h]
 	\includegraphics[width=1\textwidth]{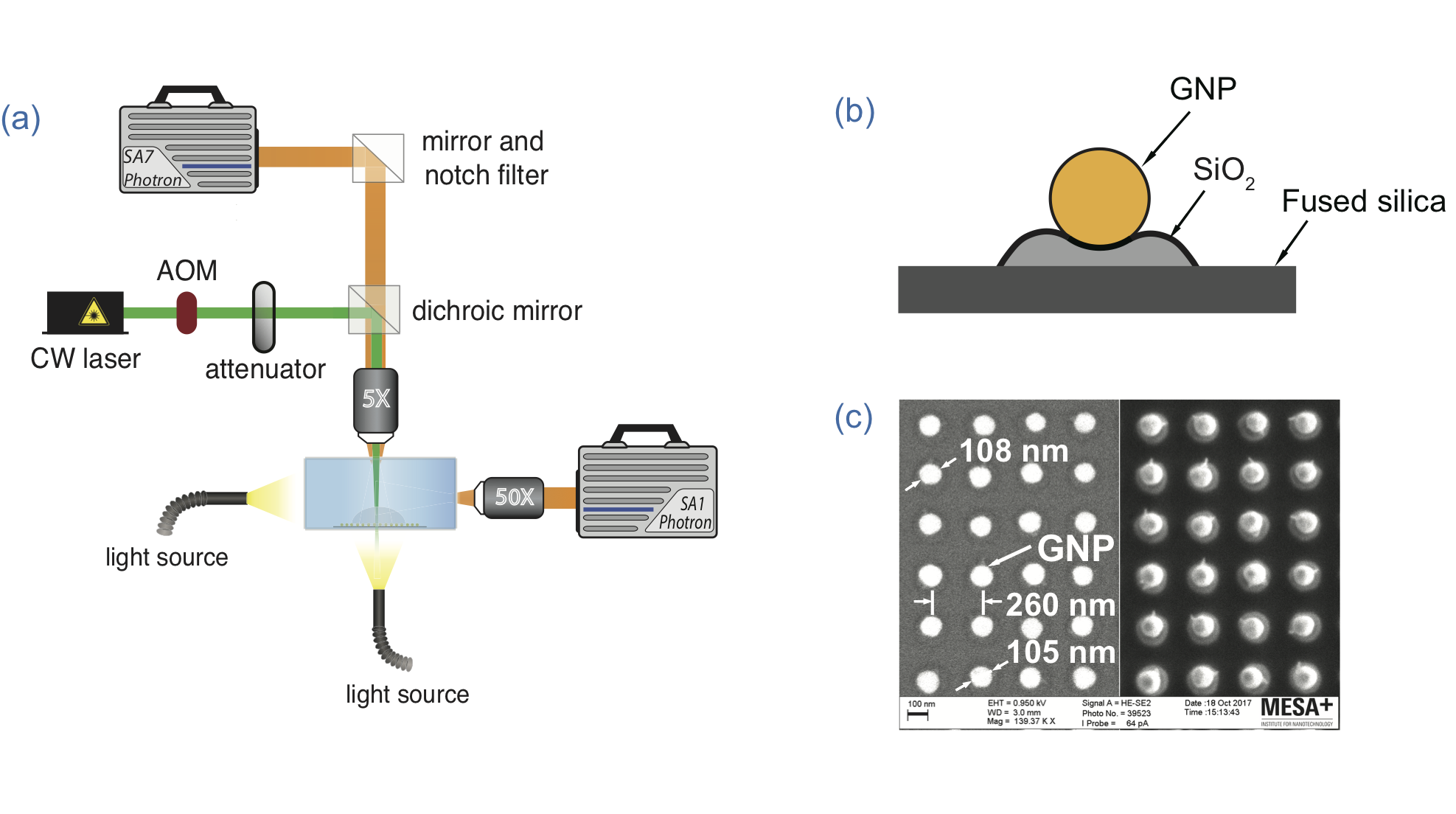}
 	\caption{
   a) Schematic of the optical imaging facilities for plasmonic microbubble formation observation, b) Schematic of a gold nanoparticle sitting on a $SiO_2$ island on a fused-silica substrate, c) SEM images of the patterned gold nanoparticle sample surface. 
 	}
 	\label{fig:experiment}
 \end{figure*}

A gold layer of approximately 45 nm was deposited on an amorphous fused-silica wafer by using an ion-beam sputtering system (home-built T$^\prime$COathy machine, MESA+ NanoLab, Twente University). A bottom anti-reflection coating (BARC) layer ($\sim$186 nm) and a photoresist (PR) layer ($\sim$200 nm) were subsequently coated on the wafer. Periodic nanocolumns with diameters of approximately 110 nm were patterned in the PR layer using displacement Talbot lithography (PhableR 100C, EULITHA) \cite{the2017}. These periodic PR nanocolumns were subsequently  transferred at wafer level to the underlying BARC layer, forming 110 nm BARC nanocolumns by using nitrogen plasma etching (home-built TEtske machine, NanoLab) at 10 mTorr and 25 W for 8 min. Using these BARC nanocolumns as a mask, the Au layer was subsequently etched by ion beam etching (Oxford i300, Oxford Instruments, United Kingdom) with 5 sccm Ar and 50-55 mA at an inclined angle of $5^{\circ}$. The etching for 9 min resulted in periodic Au nanodots supported on cone-shaped fused-silica features. The remaining BARC was stripped using oxygen plasma for 10 min (TePla 300E, PVA TePla AG, Germany). The fabricated array of Au nanodots was heated to $1100^{\circ}$ C in 90 min and subsequently cooled passively to room temperature. During the annealing process, these Au nanodots re-formed into spherical-shaped Au nanoparticles. Figure \ref{fig:experiment}b shows the schematic of a gold nanoparticle sitting on a SiO2 island on a fused-silica. The SEM image of the patterned gold nanoparticle sample surface is shown in Figure \ref{fig:experiment}c.

\subsection*{Setup description}
The experimental setup for plasmonic microbubbles imaging is shown in Figure \ref{fig:experiment}a. The gold nanoparticle covered sample was placed in a quartz glass cuvette and filled with liquid. A continuous-wave laser (Cobolt Samba) of 532 nm wavelength and a maximum power of 300 mW was used for sample irradiation. The size of the laser spot on the sample was 20 $\mu m$. An acousto-optic modulator (Opto-Electronic, AOTFncVIS) was used as a shutter to control the laser irradiation on the sample surface.  Laser pulses were generated and controlled by pulse/delay generator (BNC model 565). The laser power was controlled by using a half-wave plate and a polarizer and measured by a photodiode power sensor (S130C, ThorLabs). Two high-speed cameras were installed in the setup, one (Photron SA7) equipped with 5x long working distance objective (LMPLFLN, Olympus) and the other (Photron SA1) equipped with long working distance objective 50x (SLMPLN, Olympus) and operated at various framerates from 30 fps for the slow (second) phase of bubble shrinkage up to 300 kfps for the first (rapid) phase of bubble shrinkage. The first camera was used for a top-view and laser allignment, while the second one for the side-view. Two light sources, an Olympus ILP-1 and a Schott ACE I provided illumination for the two high-speed cameras. The optical images have been processed with an image segmentation algorithm in Python for the extraction of the bubble parameters, such as radius, height, volume and contact line. During experiments, the nanoparticle covered sample surface was immersed into deionized water (Milli-Q Advantage A10 System, Germany). Air concentration levels have been measured by an oxygen meter (Fibox 3 Trace, PreSens). For obtaining the air-equilibrated water, a sample bottle containing deionized water was kept open in air for 10 hours, the measured air concentration level is $c_g/c_{sat}$= 0.99. Partially degassed water with different air saturation levels was prepared by tuning the time of deionized water degassing in a vacuum chamber.
Under laser irradiation gold nanoparticles might be melted by the generated heat. The temperature increase of the array of Au nanoparticle can be calculated by a method developed by Baffou et al.  \cite{baffou2013_2}. Using this method, we arrive at a temperature increase of about 270 K, \textit{i.e.} far below the melting temperature of gold. In addition, our SEM images recorded after the experiments reveal that the nanoparticles are unaltered (position, size and shape).

\section{Results and discussion}

\subsection{Condensation-dominated shrinkage}

\begin{figure}[H]
 	\includegraphics[width=1\textwidth]{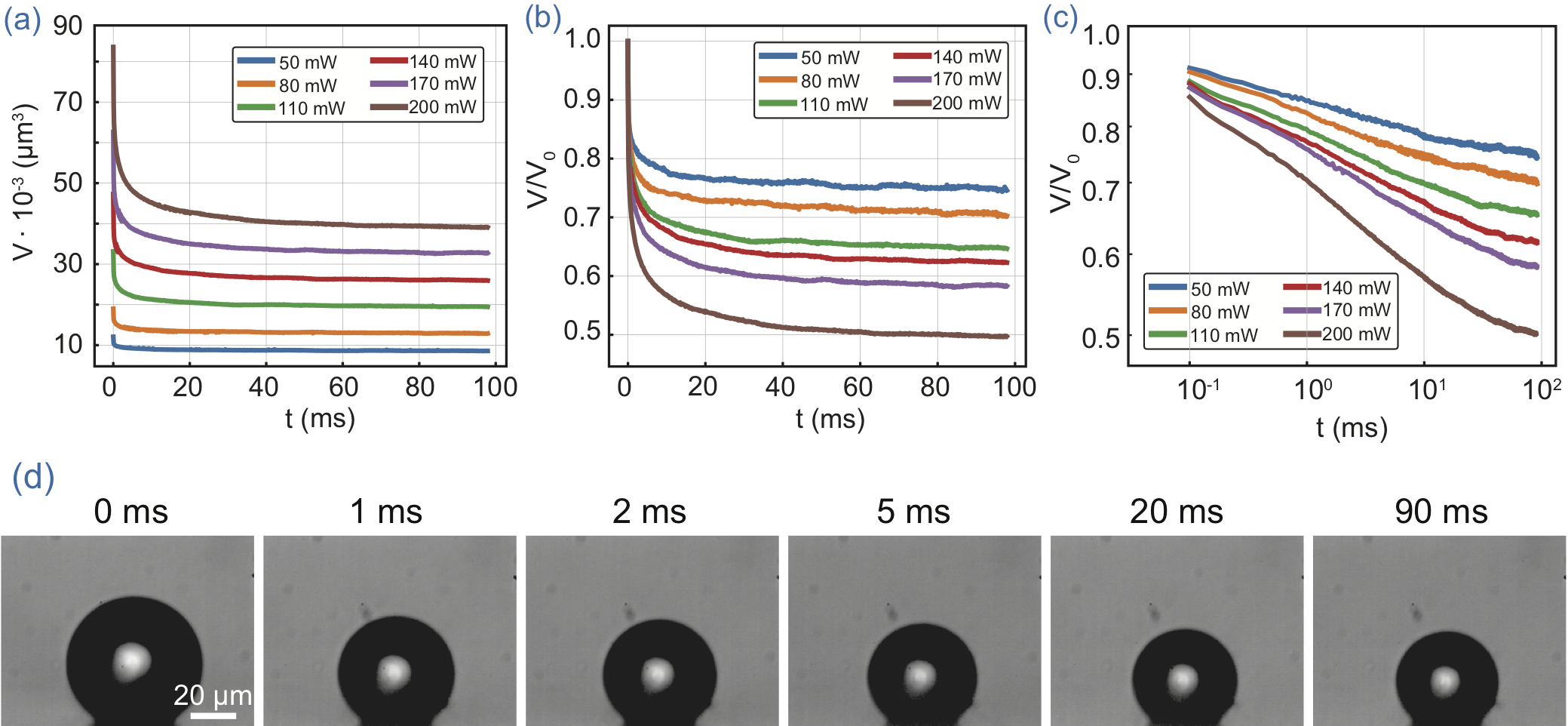}
 	\caption{ 		
 		a) Bubble volume, b) normalized bubble volume and c) normalized bubble volume in double-logarithmic scale as a function of time right after switching off the laser after 0.1 s of irradiation for various laser powers in water with $c_g/c_{sat}$ = 0.99,  d) snapshots of plasmonic bubble shrinkage in air-equilibrated water during the first 90 ms after switching off the laser after a 0.1 s laser pulse with a laser power of $P_l$ = 200 mW. The first frame (0 ms) corresponds to the moment when the laser was switched off.
 	}
 	\label{fig:rapid shrinkage example}
 \end{figure}

As was already mentioned, right after turning off the laser, the bubble quickly partially shrinks, due to water vapor condensation. To illustrate this process, the bubble volume $V$ as well as the normalized volume $V/V_0$ , where $V_0$ is bubble volume at the moment we switched the laser off, as a function of time during the first 100 ms after 0.1 s of laser irradiation in water with gas saturation level $c_g/c_{sat}$ = 0.99 for different laser powers are shown in Figures \ref{fig:rapid shrinkage example}a and \ref{fig:rapid shrinkage example}b, respectively. One can see that for all laser powers rapid bubble shrinkage occurs during the first milliseconds but then slows down, followed by much slower volume reduction. The snapshots of the process in various moments of time are shown in Figure \ref{fig:rapid shrinkage example}d. From Figure \ref{fig:rapid shrinkage example}a, it is clear that at the specified parameters $\tau_p$ = 0.1 s and $c_g/c_{sat}$ = 0.99 higher laser powers lead to larger bubbles before switching off the laser. At the same time, the initial volume drop is much larger for big bubbles than for the small ones: for the lowest laser power used in the experiments $P_l$=50 mW, the normalized bubble volume drops to 0.75 during first 100 ms after turning off the laser (Figure \ref{fig:rapid shrinkage example}b), while it reduces to 0.52 for the highest laser power $P_l$=200 mW during the same time period. This finding implies that in the second case there is relatively more water vapor in the bubble than in the first case. Indeed, with higher laser powers, more energy can be provided to the system to evaporate the liquid in the vicinity of the bubble. 

In a nutshell, for specified parameters ($\tau_p$ and $c_g/c_{sat}$) higher laser powers lead to larger bubbles with larger vapor portion in the total bubble composition and, therefore, a greater volume drop during the first milliseconds of bubble shrinkage has been observed. It is clear that the bubble behaviour during this phase is defined by the bubble composition right before turning off the laser irradiation. The previously fixed parameters  $\tau_p$  and $c_g/c_{sat}$ certainly influence the initial bubble composition as well. In order to thoroughly study these dependences we have considered a broad range of parameters: laser powers, $P_l$: 50, 80, 110, 140, 170 and 200 mW, pulse lengths $\tau_p$: 0.03, 0.05, 0.1, 0.2, 0.5 s and air saturation levels, $c_g/c_{sat}$: 0.99, 0.79, 0.64, 0.50 and 0.32.
 
The air-vapor ratio in the total bubble composition can be estimated, assuming that after the initial rapid partial shrinkage all the vapor has already condensed and considering the bubble as a pure gas bubble. This assumption is reasonable, as the complete vapor bubble condensation occurs in order of milliseconds \cite{prosperetti2017}. Therefore, the slow shrinkage part (almost horizontal part of the lines at Figure \ref{fig:rapid shrinkage example}b) can be linearly fitted and extrapolated to estimate, directly from Figure \ref{fig:rapid shrinkage example}b, the part of the bubble that has been condensed, in other words, what was the vapor portion in the initial bubble composition.

  \begin{figure}[H]
 	\includegraphics[width=0.7\textwidth]{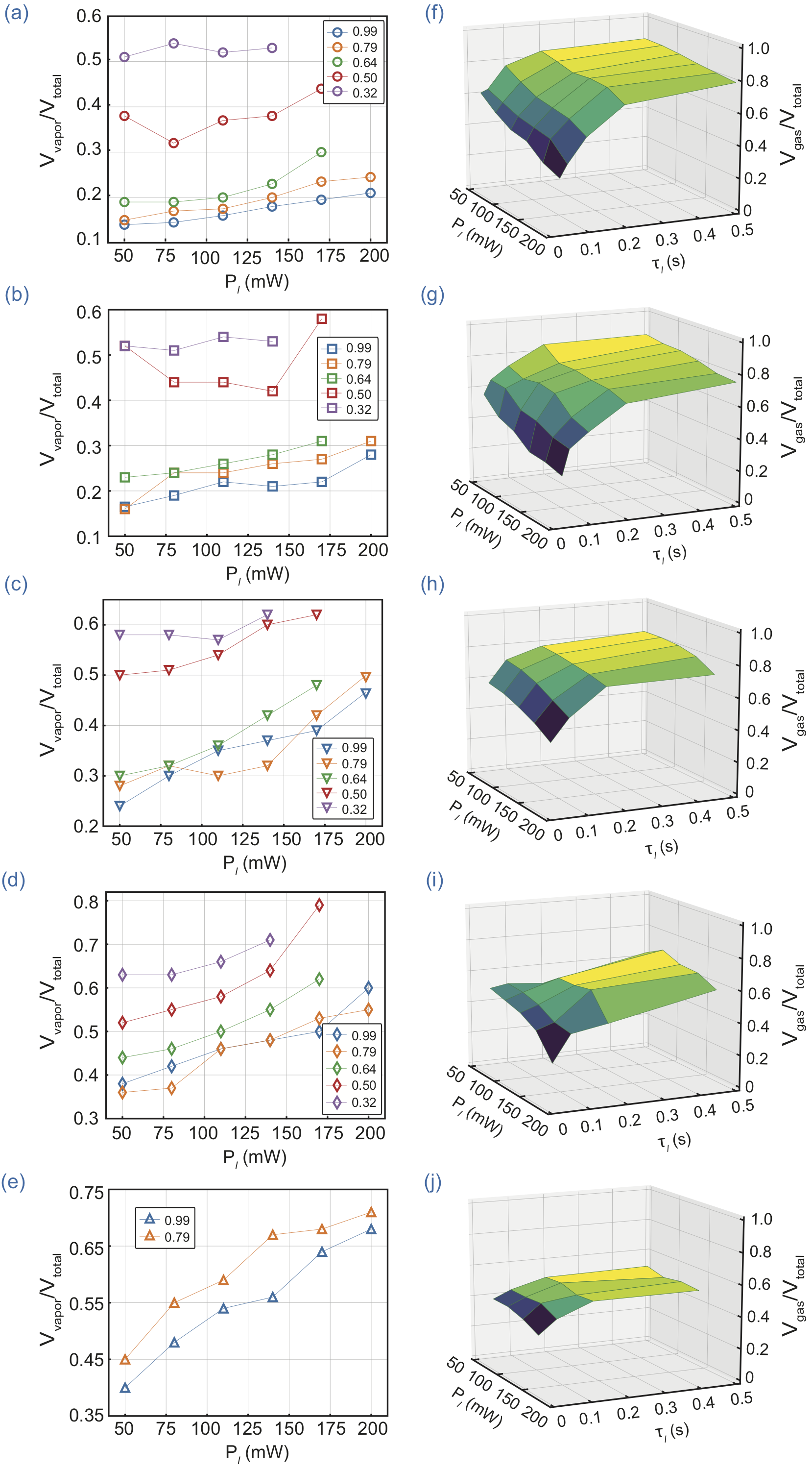}
 	\caption{
 		$V_{vapor}/V_{total}$ as a function of $P_l$ for a) $\tau_p$ = 0.5 s, b) 0.2 s, c) 0.1 s, d) 0.05 s, e) 0.03 s used for bubble formation and various water saturation levels $c_g/c_{sat}$: 0.99 (blue), 0.79 (orange), 0.64 (green), 0.50 (red), 0.32 (purple); 
 		$V_{gas}/V_{total}$ as a function of $P_l$ and $\tau_l$ for saturation levels $c_g/c_{sat}$: f) 0.99, g) 0.79, h) 0.64, i) 0.50, j) 0.32.
 	}
 	\label{fig:Vapor portion in the bubble(2)}
 \end{figure}

Following the above method water vapor portions in total bubble volume as a function of $P_l$ at various air concentrations for $\tau_l$ = 0.5, 0.2, 0.1, 0.05 and 0.03 s have been estimated and presented in figures \ref{fig:Vapor portion in the bubble(2)} (a)-(e), respectively. First of all, as expected, for all dissolved air concentrations $c_g/c_{sat}$ and laser pulse lengths $\tau_p$, the water vapor fraction $V_{vapor}/V_{total}$ increases with increasing laser power. For example, in water with $c_g/c_{sat}$ = 0.79 after 0.5 s of laser irradiation (Figure \ref{fig:Vapor portion in the bubble(2)}a) the vapor portion in the bubble composition gradually increases from 0.14 to 0.24 with increasing laser power from 50 mW to 200 mW, respectively. When more power is provided, there is more energy available in the system to evaporate the liquid during bubble growth. One can also notice that, for all pulse lengths and laser powers, bubbles, which have been formed in less saturated water, contain relatively more vapor than the ones generated in water with a higher dissolved air concentration. For example, for $\tau_l$ = 0.5 s, bubbles in water with 99\% of air concentration consist of only 0.13\dots 0.21 water vapor, whilst, approximately, half of the bubble's volume in water with 32\% of dissolved air concentration is water vapor. For less saturated water, there is lack of air to diffuse into the bubble; therefore, the bubbles are more vaporous.

The air volume portion in the initial total bubble volume as a function of $P_l$ and $\tau_p$ for air saturation levels $c_g/c_{sat}$  = 0.99, 0.79, 0.64, 0.50 and 0.32, respectively, is shown in figures \ref{fig:Vapor portion in the bubble(2)} (f)-(j). The air fraction in the bubble rapidly decreases with shortening of the laser pulse, which means that bubbles become more vaporous. A similar trend was observed for all water saturation levels. For example, in water with air concentration of 99\% (Figure \ref{fig:Vapor portion in the bubble(2)}f) the air portion in the bubble composition, under the lowest laser power in our experiments - $P_l$ = 50 mW, decreases from 0.87 to 0.59 with decreasing laser pulse duration from 0.5 s to 0.03 s, and decreases from 0.79 to 0.31 for the highest laser power $P_l$ = 200 mW. The plasmonic microbubble during its steady growth undergoes two consecutive phases: an initial phase, governed by liquid evaporation ($\sim$ 10...20 ms) and a second phase, governed by air diffusion from the bulk liquid ($>$0.1 s) \cite{wang2017}. Therefore, for short laser pulses bubble growth is mainly in the evaporation-controlled regime and there is insufficient time for air to diffuse into the bubble. Therefore, these plasmonic bubbles are more vaporous. With increasing pulse duration, the transition to the diffusive-controlled regime occurs and, therefore, these bubbles turn into more gaseous bubbles. It is worth while to mention that even for the short laser pulses ($\tau_p = 0.03$), during which bubble growth is mainly governed by water evaporation, a considerable amount of air is still present in the bubble, reaching more than a half of the bubble volume for low laser powers.

\begin{figure}[H]
 	\includegraphics[width=0.9\textwidth]{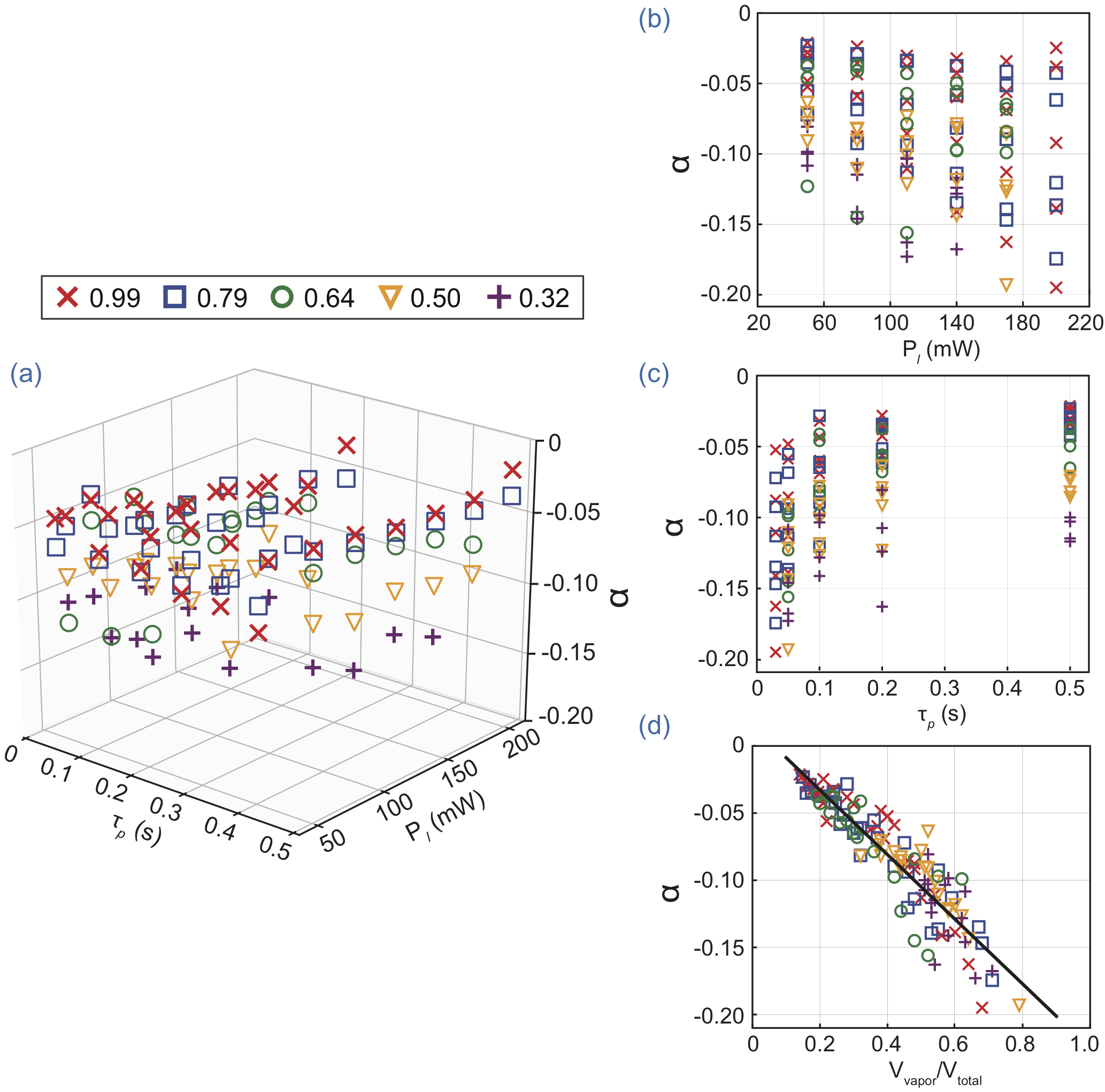}
 	\caption{
 		a) The effective scaling exponent $\alpha$ in the $V(t) \propto t^\alpha$ during the first 5 ms of bubble shrinkage as a function of the pulse length $\tau_p$ and laser power $P_l$, b) projection to the $\alpha - P_l$ plane, c) projection to the $\alpha - \tau_p$ plane, d) the effective scaling exponent $\alpha$ as a function of vapour portion in initial total bubble composition for various air saturations $c_g/c_{sat}$: 0.99 (red cross),  0.79 (blue square), 0.64 (green circle), 0.50 (yellow triangle) and 0.32 (purple cross). For all concentrations the same dependence of decreasing $\alpha$ with increasing vapour ratio is observed.
 	}
 	\label{fig:scaling exp 1st phase}
 \end{figure}

In order to estimate the influence of the initial bubble composition on the first shrinkage phase, one can extract the effective scaling exponent $\alpha$ from $V(t)\propto t^\alpha$ for the first 5 milliseconds after the end of laser irradiation. The effective scaling exponent as a function of $P_l$ and $\tau_p$ is shown in Figure \ref{fig:scaling exp 1st phase}a with its projections to the $\alpha - P_l$ plane and $\alpha - \tau_p$ plane in Figures \ref{fig:scaling exp 1st phase}b and \ref{fig:scaling exp 1st phase}c, respectively. No clear dependence can be figured out from these plots. However, it turned out that $\alpha$ strongly depends on the water vapor portion in the initial bubble composition; a clear linear trend can be observed in Figure \ref{fig:scaling exp 1st phase}d. The more air is contained in the bubble, the smaller the absolute value of $\alpha$ is. The behaviour is universal for all considered $P_l$, $\tau_p$ and $c_g/c_{sat}$.

\subsection{Comparison with Rayleigh-Plesset type model and the diffusion equation}

For a more quantitative description of the process, we compare our experimental results with a model developed for bubble growth/shrinkage, which takes into account temperature and air diffusion in the liquid and temperature and air-vapor diffusion in the bubble.\cite{hao2017} The model approximates the bubble as a complete sphere. Both the gas and the vapor are assumed to behave like ideal gases. The radial motion of the bubble is described by the Rayleigh-Plesset equation corrected for compressibility effects of the liquid:
\begin{equation}
 \left(1-{\dot{R}\over c_L}\right) \, R\ddot{R}+{3\over 2} \left(
1-{\dot{R}\over 3 c_L}\right) \dot{R}^2  \, = \, {1\over \rho_L}
\left(1+{\dot{R}\over c_L}+{R\over c_L}{d\over dt}\right) \,
\left( p -P- {2\sigma \over R} - 4 \mu_L {\dot{R}\over R} \right) \,,
\label{rp}
\end{equation}

\noindent where $R$ is the bubble radius, $p$ the pressure in the bubble, $P$ the ambient pressure, $\sigma$ the interfacial tension coefficient, and $\rho_L$, $\mu_L$ and $c_L$ the liquid density, viscosity and speed of sound; dots denote time differentiation and the subscript $L$ liquid quantities.

\def\pmb#1{\setbox0=\hbox{#1}
        \kern-.025em\copy0\kern-\wd0
        \kern.05em\copy0\kern-\wd0
        \kern-.025em\raise.0433em\box0 } 
\def\div{\pmb {$\nabla$}}

The temperature field in the air-vapor mixture is obtained from the enthalpy equation including the effect of air-vapor diffusion:
\begin{equation}
\frac{\gamma_V}{\gamma_V-1}\left(1- \frac{\gamma_G-\gamma_V}{\gamma_V
(\gamma_G-1)} \frac{p_G}{p}\right)\frac{p}{T}\frac{dT}{dt}\,=\, \dot{p} 
+\div \cdot(k\div T)+ (c_{pV}-c_{pG}) \rho {\cal D} (\div C\cdot \div T)\,,
\label{velfi}
\end{equation}

\noindent where $k$ is the thermal conductivity of the air-vapor mixture, $c_p$ the specific heat at constant pressure and  $\gamma$ the ratio of specific heats; indices $V$ and $G$ refer to the vapor and gas, respectively. Furthermore, $T$ is the absolute temperature, $dT/dt$ the convective derivative and $C$ the air concentration by mass. Both the air-vapor velocity field and the internal bubble pressure (assumed spatially uniform) are determined from a combination of the continuity and energy equations.\cite{hao2017}

Conservation of the gas, liquid/vapor and energy across the bubble interface is imposed. The model is solved numerically by a spectral method which reduces the problem to a system of ordinary differential equations. In particular, the temperature and concentration fields in the liquid are expressed in a Chebyshev polynomial series in the variable $x=R(t)/r$ with $r$ the 
distance from the bubble center. For a more detailed description the reader is referred to a recent publication.\cite{hao2017}

The calculation starts at the instant in which the laser is turned off. The measured bubble radius is used as the initial radius. The initial temperature distribution in the liquid is approximated  by a fourth-order polynomial even in $x$. The three coefficients of the linear combination are determined from the ambient liquid temperature, the vanishing of the heat flux at $x=0$ (i.e., at $r\rightarrow \infty$), and the surface bubble temperature $T_s$, used as a fitting parameter. The initial air concentration in the 
liquid is approximated in the same way. Given a provisional value for $T_s$, with the assumption of thermodynamic equilibrium, the partial pressure of the vapor is known, and the partial pressure of air follows from the assumption of mechanical equilibrium. 
Through Henry's law, this determines the air concetration at the bubble surface. The air-vapor mixture is assumed to be initially homogeneous at the assumed initial temperature. The results are somewhat dependent on this initial distribution, which cannot be accurately determined from the experiment.

\begin{figure}[H]
 	\includegraphics[width=1.0\textwidth]{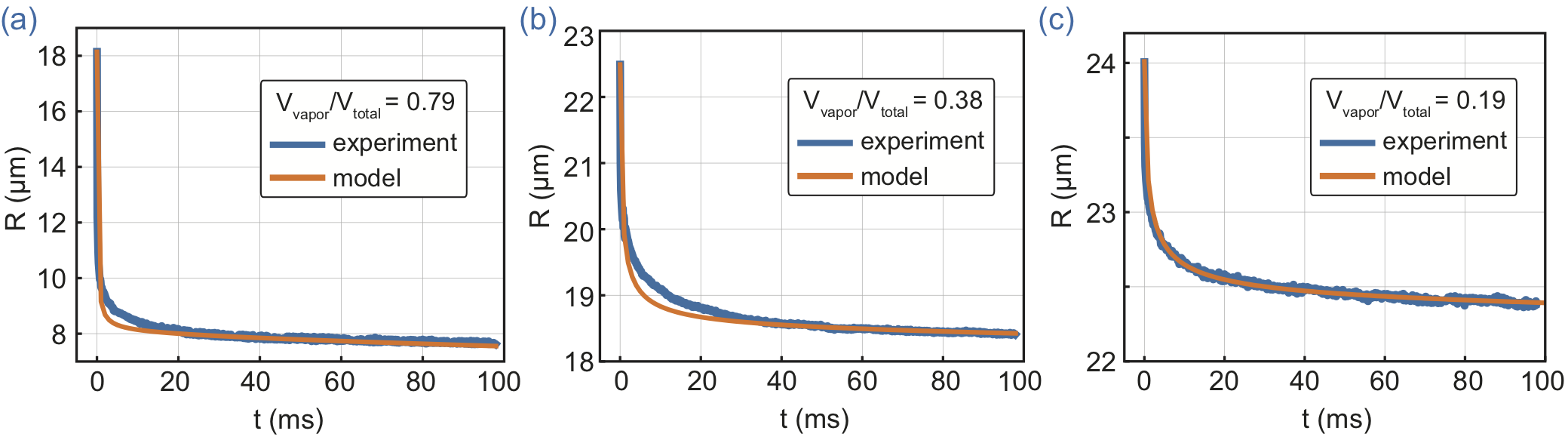}
 	\caption{
		Comparison of the experimental results (blue) and model predictions (orange) of the bubble radius change with time during first 100 ms after switching off the laser for various initial conditions: a) $P_l$ = 170 mW, $\tau_p$ = 0.05 and $c_g/c_{sat}$ = 0.5, $V_{vapor}/V_{total}$ = 0.79; b) $P_l$ = 140 mW, $\tau_p$ = 0.1 and $c_g/c_{sat}$ = 0.99, $V_{vapor}/V_{total}$ = 0.38; c) $P_l$ = 50 mW, $\tau_p$ = 0.5 and $c_g/c_{sat}$ = 0.99, $V_{vapor}/V_{total}$ = 0.19
 	}
 	\label{fig:exp vs model}
 \end{figure}

In spite of this uncertainty, it can be seen from Figure \ref{fig:exp vs model} that the model is able to reproduce the bubble evolution reasonably well after the laser has been switched off for as long the data were recorded. The error is noticeable during the time in which the bubble shrinkage switches from condensation dominated to diffusion dominated regime, which is when the influence of the unknown initial temperature distribution in the liquid is particularly significant. Indeed, in Figure \ref{fig:exp vs model}c, for which the laser power is small and the liquid is nearly saturated with air so that the initial amount of vapor in the bubble is small, the agreement is significantly better than in the other cases. We have found a similar level of agreement for all the dissolved air concentrations that we have tested. 

To summarize, the bubble dynamics during the first phase depends on the initial bubble composition, which is defined by the conditions ($P_l$, $\tau_p$, $c_g/c_{sat}$) during the bubble growth, i.e. on the history of bubble formation, which we can quantitatively describe with the model of ref. [42].

\subsection{Diffusion-dominated shrinkage}

\begin{figure}
 	\includegraphics[width=0.9\textwidth]{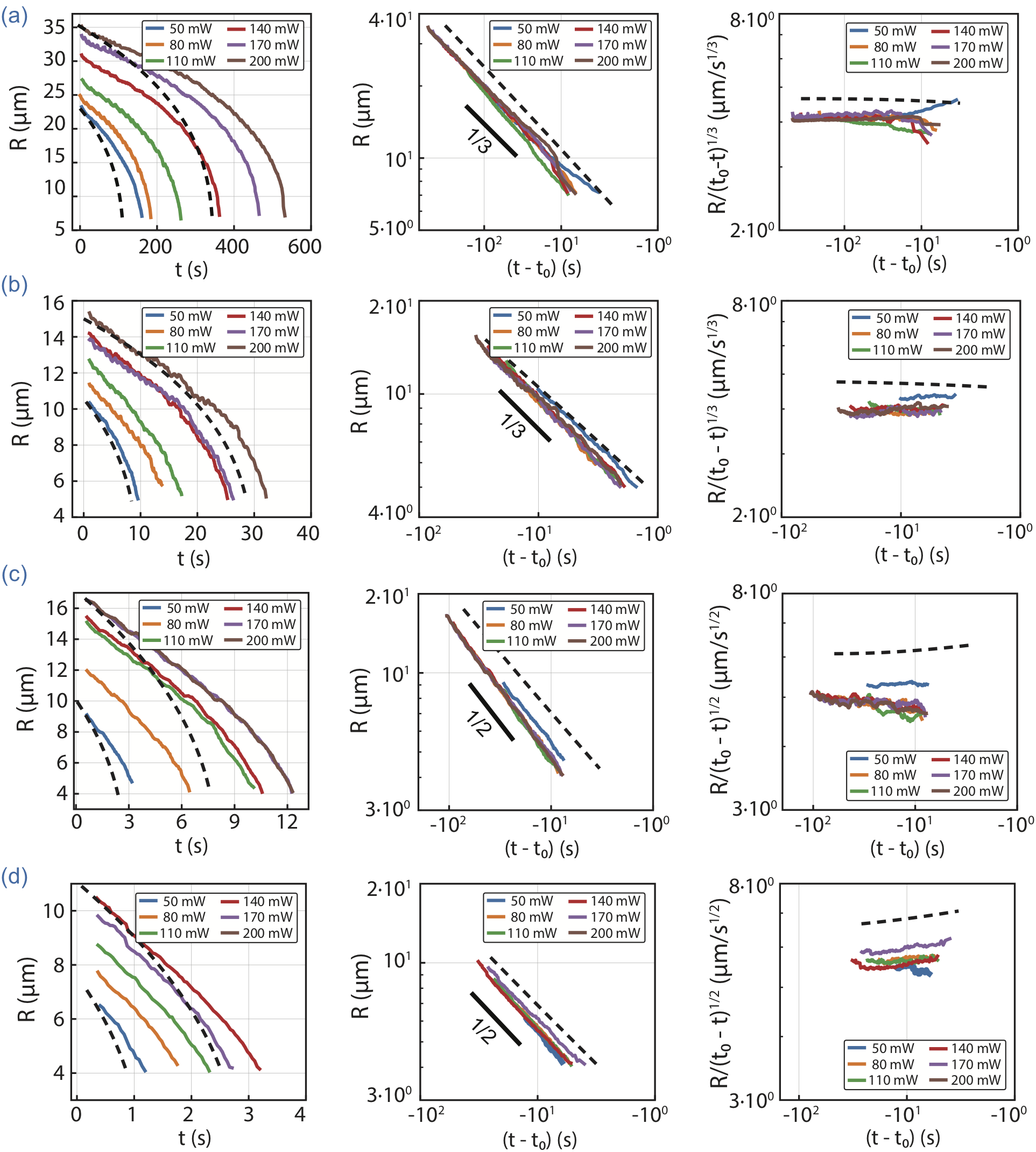}
 	\caption{
 		Bubble radius $R$ as a function of time $t$ for various experimental conditions:  a) $c_g/c_{sat}$ = 0.99, $\tau_p$ = 0.5 s, b) $c_g/c_{sat}$ = 0.99, $\tau_p$ = 0.05 s, c) $c_g/c_{sat}$ = 0.50, $\tau_p$ = 0.5 s, d) $c_g/c_{sat}$=0.99, $\tau_p$=0.05 s and various laser powers $P_l$: 50 mW (blue), 80 mW (orange), 110 mW (green), 140 mW (red), 170 mW (purple) and 200 mW (brown). The second and third columns represent $R$ as a function of $t$ in double logarithmic scale and compensated plots $R/(t_0 - t)^{1/3}$ (a,b) and $R/(t_0 - t)^{1/2}$ (c,d) as functions of time in double logarithmic scale, respectively. Black dashed lines are theoretical curves obtained from the direct solution of the purely diffusive bubble dissolution problem, see text.
 	}
 	\label{fig:2nd phase analysis}
 \end{figure}

After the rapid bubble shrinkage, a much slower process takes over. During this phase the bubble dissolves due to air diffusion from the bubble to the bulk liquid. Our experimental results for laser powers $P_l$ = 50 mW, 80 mW, 110 mW, 140 mW, 170 mW and 200 mW, pulse durations $\tau_p$ = 0.5 s and 0.05 s and water with air saturation levels $c_g/c_{sat}$ = 0.99 and 0.50 are summarized in Figure \ref{fig:2nd phase analysis}. In Figure \ref{fig:2nd phase analysis}a the bubble radii $R$ as function of time $t$ for $c_g/c_{sat}$ = 0.99 and $\tau_p$ = 0.5 s along with a double logarithmic plot and a double logarithmic plot of $R$ compensated with $(t_0 - t)^{1/3}$ are shown, where $t_0$ is the lifetime of the bubble. Figures \ref{fig:2nd phase analysis}(b)-(d) represent the same set of figures, but for different experimental conditions: $c_g/c_{sat}$ = 0.99 and $\tau_p$ = 0.05 s, $c_g/c_{sat}$ = 0.50 and $\tau_p$ = 0.5 s, $c_g/c_{sat}$ = 0.50 and $\tau_p$ = 0.05 s, respectively.

As was already mentioned, higher laser powers and longer laser pulses lead to larger bubbles and, therefore, at the same ambient conditions, it takes longer for them to completely dissolve. For example, in water with 99\% of dissolved air concentration it takes around 100 s for the bubble, which was formed after laser pulse with $\tau_p$ = 0.5 s and $P_l$ = 50 mW, and initial radius $R_0 = 23\ \mu m$ to dissolve, while it takes around 550 s for the one, formed with $P_l$ = 200 mW, and with $R_0 = 36\ \mu m$. Bubbles of comparable size dissolve much faster in degassed water than in saturated water, \textit{e.g.} the bubble with $R= 15\ \mu m$ completely dissolves in water with $c_g/c_{sat}$ = 0.99 in around 33 s (Figure \ref{fig:2nd phase analysis}b), while in water with $c_g/c_{sat}$ = 0.50 in 10 s (Figure \ref{fig:2nd phase analysis}c), due to the larger concentration gradients between the bubble and the bulk liquid.

The solid black lines in the double logarithmic plots correspond to the effective scaling exponent of ${1/3}$ (Figures \ref{fig:2nd phase analysis}a, \ref{fig:2nd phase analysis}b) and ${1/2}$ (Figures \ref{fig:2nd phase analysis}c,  \ref{fig:2nd phase analysis}d) in $R(t)\propto (t_0 - t)^\alpha$, and are shown for clearer indication of the slope. One can see that not only for all laser powers, but also for all pulse durations bubbles follow similar dynamics. The double logarithmic plots of $R$ compensated by $(t_0 - t)^{1/3}$ (Figures \ref{fig:2nd phase analysis}a,  \ref{fig:2nd phase analysis}b) and $(t_0 - t)^{1/2}$ (Figures \ref{fig:2nd phase analysis}c, \ref{fig:2nd phase analysis}d) further confirm the value of the effective scaling exponent. Despite the slight variations, the scaling exponent remains close to $1/3$ for $c_g/c_{sat}$ = 0.99 and close to $1/2$ for $c_g/c_{sat}$ = 0.50 for all experimental parameters with fixed air concentration.

\begin{figure}[H]
 	\includegraphics[width=0.5\textwidth]{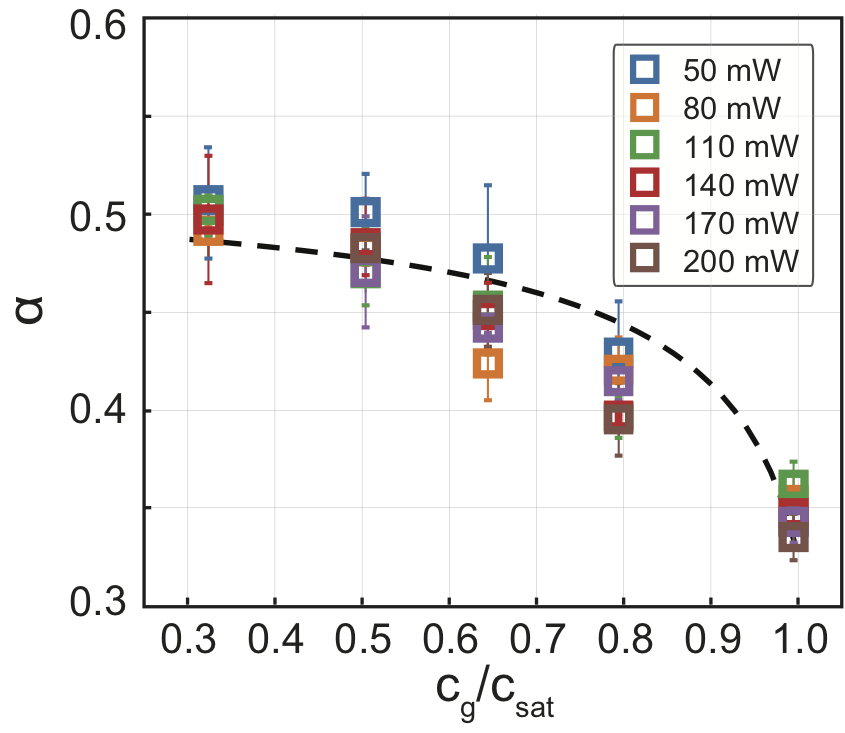}
 	\caption{
		Effective scaling exponent $\alpha$ in $R(t) \propto (t_0 - t)^\alpha$ as a function of air concentration level for various laser powers $P_l$: 50 mW (blue), 80 mW (orange), 110 mW (green), 140 mW (red),170 mW (purple) and 200 mW (brown). The dashed black line is the theoretical curve obtained from a complete solution of equation (\ref{main_eq}) for various air concentrations, showing good agreement with our data.
 	}
 	\label{fig:scaling exp 2nd phase}
 \end{figure}

In order to further investigate how the concentration of air dissolved in liquid influences the long-term bubble dissolution dynamics, we performed a series of experiments with various $c_g/c_{sat}$, namely, $c_g/c_{sat}$ = 0.99, 0.79, 0.64, 0.50 and 0.32 and $\tau_p$ = 0.5 s, 0.2 s, 0.1 s, 0.05 s, 0.03 s. Performing the same analysis as discussed above for all data sets, the effective scaling exponents $\alpha$ in $R(t) \propto (t_0 - t)^\alpha$ for various saturation levels have been extracted (see Figures S1, S2, S3, and S4 in Supplementary Information). Remarkably, $\alpha$ only slightly varies with $P_l$ and $\tau_p$, however it strongly depends on the concentration of dissolved air in the liquid. For water with $c_g/c_{sat}$ = 0.99 almost irrespective of $P_l$ and $\tau_p$ the average effective scaling exponent is always around 0.34. For water degassed till 79\% of the saturation air concentration $\alpha$ increases to 0.41. Further liquid degassing leads to larger scaling exponents, namely, for $f_g=0.64$ it increases to $\alpha=0.45$, for $c_g/c_{sat}=0.50$ to $\alpha=0.48$ and for $c_g/c_{sat}=0.32$ to $\alpha=0.50$ (Figure \ref{fig:scaling exp 2nd phase}).
So what defines the scaling exponent? 

\begin{figure}
 	\includegraphics[width=1\textwidth]{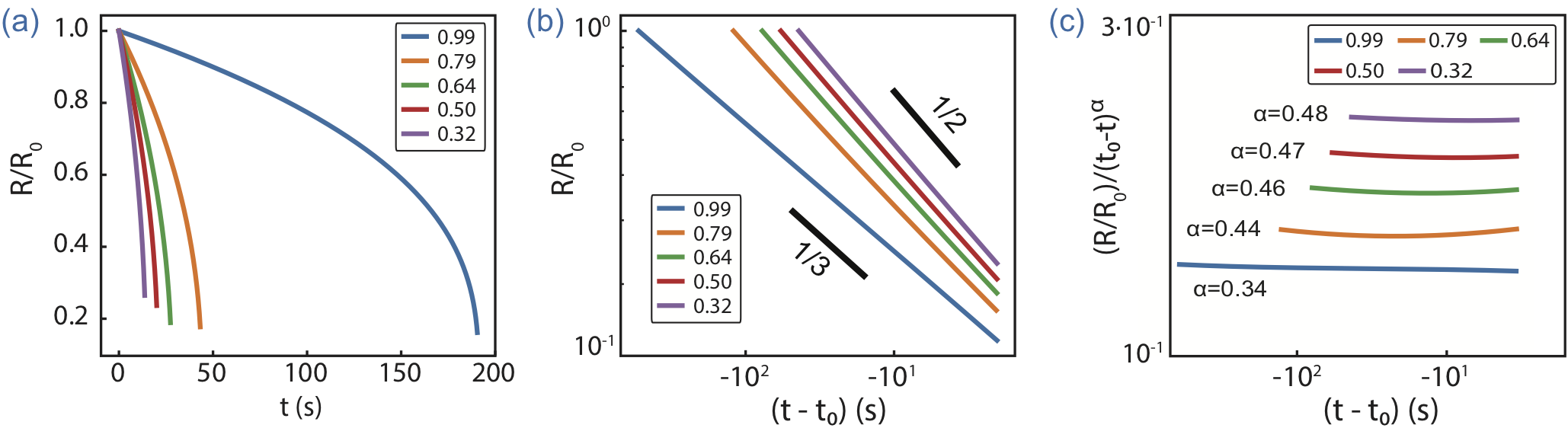}
 	\caption{
		a) Normalized bubble radius $R/R_0$ as a function of time $t$, derived from the solution of equations (\ref{main_eq}), where $R_0 = 30 \mu m$, b) Normalized bubble radius $R/R_0$ as a function of time $t$ in double logarithmic scale and c) compensated plots $(R/R_0)/(t_0 - t)^{\alpha}$ in double logarithmic scale, where $\alpha$ is an effective scaling exponent, for various air saturation levels $c_g/c_{sat}$: 0.99 (blue), 0.79 (orange), 0.64 (green), 0.50 (red) and 0.32 (purple).
 	}
 	\label{fig:EP_solutions}
 \end{figure}
 
We consider a spherical purely gas bubble, resting in the surrounding liquid \cite{epstein_plesset}. Assuming the gas inside the bubble as ideal, \textit{i.e.} $PV = N \Re T$, and taking into account that the bubble inner pressure is given as a sum of ambient pressure and the Laplace pressure, the loss rate of the number $N$ of molecules in the bubble can be derived: 

\begin{equation}
	 \dot N  = \dfrac{4\pi \dot R R^2}{\Re T} \bigg( P_\infty +\dfrac{4}{3}\dfrac{\sigma}{R} \bigg) , 
	 \label{molec_variation1}
\end{equation}

\noindent where $R$ is the bubble radius, $\Re$ the ideal gas constant, $T$ the bubble temperature, $P_\infty$ the ambient pressure, and $\sigma$ the surface tension of the gas-liquid interface.

On the side of the liquid the molecular flux of air molecules due to the diffusion through the bubble interface is determined by the gas concentration gradient,

\begin{equation}
	 \dot N  = \left. 4 \pi R^2 D \dfrac{\partial c}{\partial r}  \right|_{r=R}, 
	 \label{molec_variation2}
\end{equation}

\noindent where $D$ is the mass diffusion coefficient and $c(r,t)$ is air concentration.

\noindent Combining equations (\ref{molec_variation1}) and (\ref{molec_variation2}) we have

\begin{equation}
	 \dot R  \bigg(1 +\dfrac{4}{3}\dfrac{\sigma}{R P_\infty} \bigg) \dfrac{P_\infty}{\Re T}   = \left. D \dfrac{\partial c}{\partial r}  \right|_{r=R}.
	 \label{R_dot}
\end{equation}

\noindent The gas concentration field within the liquid is given by the convection-diffusion equation

\begin{equation}
	\partial_t(c(r,t)) = \frac{D}{r^2}\partial_r(r^2 \partial_r c(r,t)) + \frac{a^2 \dot{a}}{r^2} \partial_r c(r,t).
	\label{gas_diffusion}
\end{equation}

\noindent For the majority of the diffusion process, we can make a quasi-steady approximation to the gradient of the dissolved gas concentration at the bubble surface and neglect the time-dependent terms to write

\begin{equation}
	\left. \partial_r c_g \right|_{r=R} \simeq \frac{c_\infty - \left. c_g \right|_{r=R}}{R}.
	\label{approx}
\end{equation}

\noindent The dissolved air concentration at the bubble interface $c_g$ will be related to the saturated air concentration under a plane interface $c_{sat}$ according to Henry's law as

\begin{equation}
	\left. c_g \right|_{r=R} = c_{sat} (P_\infty)  \bigg(1 + \frac{2\sigma}{RP_\infty} \bigg).
	\label{c_interface}
\end{equation}

\noindent Combining ({\ref{R_dot}}), (\ref{approx}) and (\ref{c_interface}) and introducing $\zeta = 1 - {c_\infty}/{c_{sat}(P_\infty)}$ and $\xi = c_{sat}(P_\infty) \Re T / P_\infty$ one can finally obtain:

\begin{equation}
	\bigg(1 +\dfrac{4}{3}\dfrac{\sigma}{R P_\infty} \bigg) R \dot R = - D \xi \bigg(\zeta +\frac{2\sigma}{RP_\infty} \bigg).
	\label{main_eq}
\end{equation}

\noindent If one neglects surface tension effects, the equation reduces to

\begin{equation}
	R \dot R  = - D \xi \zeta,
	\label{no_surface}
\end{equation}

\noindent and can be easily integrated to find $R(t) \propto (t_0 - t)^{1/2}$. Indeed, in our experiments the effective scaling exponent $\alpha$ was found to be close to $1/2$ for degassed water with $c_g/c_{sat} \leq 0.5$.

The other limiting case is the case of fully saturated liquid, so no concentration gradient exists in the system, \textit{i.e.} $\zeta = 0$.  Omitting the second term in the sum on the left in equation \ref{main_eq}, which is generally a small number for bubbles larger than $5 \mu m$, and setting $\zeta = 0$ we have

\begin{equation}
	 R \dot R = - D \xi \frac{2\sigma}{RP_\infty},
	\label{no_gradient}
\end{equation}

\noindent from which after integration one immediately finds the scaling law $R(t) \propto (t_0 - t)^{1/3}$, which is consistent with our experiments for water with $c_g/c_{sat}=0.99$.

\noindent The complete solution of equation \ref{main_eq} can also be readily find by integration:

\begin{equation}
	1-\frac{R^2}{{R_0}^2} - 2\bigg(1 - \frac{2}{3}\zeta \bigg) \frac{\Sigma}{\zeta}
	\bigg[1 - \frac{R}{R_0} + \frac{\Sigma}{\zeta}\ln\big(\frac{\Sigma/\zeta+R/R_0}{\Sigma/\zeta+1}\big) \bigg] = 2\zeta \xi \frac{Dt}{{R_0}^2},
	\label{full_solution}
\end{equation}

\noindent where $\Sigma = 2\sigma/R_0P_\infty$.

It is reasonable to assume an effective scaling law for the bubble radius $R(t) \propto (t_0-t) ^ {\alpha(c_g/c_{sat})}$, where the effective exponent $\alpha = f(c_g/c_{sat})$ is a function of the dissolved air concentration. The normalized bubble radius $R/R_0$, where $R_0 = 30\ \mu m$ as a function of time $t$, derived from the analytical solution (\ref{full_solution}) is shown in Figure \ref{fig:EP_solutions}. One can then extract $\alpha$ from the double logarithmic plot (Figure \ref{fig:EP_solutions}b) using a linear fit. The derived effective exponents are confirmed by the double logarithmic plot compensated by $(t_0 - t)^{\alpha}$ (Figure \ref{fig:EP_solutions}c), where $\alpha$=0.34, 0.44, 0.46, 0.47 and 0.48 for $c_g/c_{sat}$=0.99, 0.79, 0.64, 0.50 and 0.32, respectively. 

The dependence of $\alpha$ on $c_g/c_{sat}$, resulting from the complete solution (\ref{full_solution}) for various air saturation levels, is shown as dashed black line in Figure \ref{fig:scaling exp 2nd phase}. Just as in our experiments with decreasing air concentration, one can see the gradual transition from the limiting regime, where water is almost fully saturated with air and the effective scaling exponent is $1/3$, to the regime where we are dealing with large concentration gradients due to water degassing, and the $1/2$ effective scaling law shows up.

\section{Conclusion}

The dynamics of plasmonic microbubble shrinkage has been studied. During the dissolution a bubble undergoes two different phases. First, rapid bubble shrinkage due to water vapor condensation occurs in the first milliseconds after switching off the laser irradiation. The dynamics of this phase highly depends on the history of bubble formation, which defines the initial bubble composition. The bubble contains more vapor in its composition and, therefore, shrinks faster for shorter laser pulse durations, lower dissolved air concentrations and higher laser powers. Afterwards, the second phase of slow bubble dissolution takes over, which is governed by air diffusion from the bubble to the bulk liquid and may last up to minutes or even hours. The dynamics of the plasmoninc bubble during this phase barely depends on the laser power and laser pulse duration, used to form the bubble, but strongly depends on the air saturation level in the bulk water, as it defines the concentration field in the liquid. The gradual transition from the effective scaling exponent $\alpha=1/3$ for almost fully saturated water with $c_g/c_{sat}=0.99$ to $\alpha\approx 1/2$ for degassed water with $c_g/c_{sat}\leq0.5$ has been demonstrated both experimentally and theoretically.

\begin{suppinfo}

Experimental results on the diffusion-dominated phase for water with $c_g/c_{sat}$ = 0.99, 0.79, 0.64, 0.50 and 0.32 for various pulse lengths and laser powers.

\end{suppinfo}

\begin{acknowledgement}

{The authors thank Hai Le The for the sample preparation and Xiaojue Zhu for fruitful discussions on numerical simulations. The authors also thank Dutch Organization for Research (NWO), Netherlands Organisation for Applied Scientific Research (TNO) and the Netherlands Center for Multiscale Catalytic Energy Conversion (MCEC) for financial support.}

\end{acknowledgement}

\bibliography{shrinkage}

\end{document}